\documentclass[conference]{IEEEtran}
\usepackage[OT1]{fontenc} 
\IEEEoverridecommandlockouts
\usepackage{cite}
\usepackage{amsmath,amssymb,amsfonts}
\usepackage{algorithmic}
\usepackage[linesnumbered,ruled]{algorithm2e}
\usepackage{graphicx}
\usepackage{textcomp}
\usepackage{xcolor}
\def\BibTeX{{\rm B\kern-.05em{\sc i\kern-.025em b}\kern-.08em
    T\kern-.1667em\lower.7ex\hbox{E}\kern-.125emX}}
\begin{document}

\title{Population-Based Hierarchical Non-negative Matrix Factorization for Survey Data\\
\thanks{This work was partially supported by NSF grants DMS-2011140, DMS-2108479, and DMS-2103093.}
}

\author{\IEEEauthorblockN{Xiaofu Ding}
\IEEEauthorblockA{\textit{Math Dept.,}
\textit{UCLA}\\
Los Angeles, United States \\
jding62@g.ucla.edu}
\and
\IEEEauthorblockN{Xinyu Dong}
\IEEEauthorblockA{\textit{Math Dept.,}
\textit{UCLA}\\
Los Angeles, United States \\
xinyudong@g.ucla.edu}
\and
\IEEEauthorblockN{Olivia McGough}
\IEEEauthorblockA{\textit{Math Dept.,}
\textit{Reed College}\\
Portland, United States \\
mcgougho@reed.edu}
\and
\IEEEauthorblockN{Chenxin Shen}
\IEEEauthorblockA{\textit{Math Dept.,}
\textit{UCLA}\\
Los Angeles, United States \\
amyshen@g.ucla.edu}
\and
\IEEEauthorblockN{Annie Ulichney}
\IEEEauthorblockA{\textit{Appl. Math Dept.,}
\textit{Yale University }\\
New Haven, United States \\
annie.ulichney@yale.edu}
\and
\IEEEauthorblockN{Ruiyao Xu}
\IEEEauthorblockA{\textit{Math Dept., UCLA} \\
Los Angeles, United States \\
ruiyaox@g.ucla.edu}
\and
\IEEEauthorblockN{William Swartworth}
\IEEEauthorblockA{\textit{Math Dept., UCLA} \\
Los Angeles, United States \\
wswartworth@gmail.com}
\and
\IEEEauthorblockN{Jocelyn T. Chi}
\IEEEauthorblockA{\textit{Math Dept., UCLA} \\
Los Angeles, United States \\
jtchi@math.ucla.edu}
\and
\IEEEauthorblockN{Deanna Needell}
\IEEEauthorblockA{\textit{Math Dept., UCLA} \\
Los Angeles, United States \\
deanna@math.ucla.edu}
}

\maketitle

\thispagestyle{plain}
\pagestyle{plain}

\begin{abstract}
Motivated by the problem of identifying potential hierarchical population structure on modern survey data containing a wide range of complex data types, we introduce population-based hierarchical non-negative matrix factorization (PHNMF). PHNMF is a variant of hierarchical non-negative matrix factorization based on feature similarity.  As such, it enables an automatic and interpretable approach for identifying and understanding hierarchical structure in a data matrix constructed from a wide range of data types.  Our numerical experiments on synthetic and real survey data demonstrate that PHNMF can recover latent hierarchical population structure in complex data with high accuracy.  Moreover, the recovered subpopulation structure is meaningful and can be useful for improving downstream inference.
\end{abstract}


\section{Introduction}
\label{sec:intro}

Modern survey data are rapidly growing larger and more complex.  Technological improvements in recent years and the ubiquity of social media have dramatically lowered barriers to collecting and storing a wide range of data types on numerous respondents and variables.  Indeed, modern survey data now frequently contain not only traditional multiple choice questions but also a wide variety of questions enabling complex, open-form short responses.  

A key issue in many studies involving survey data is the identification of latent hierarchical groups, or subpopulations, within the observations.  However, traditional statistical methods for identifying subpopulations in survey data were developed prior to modern technologies that enable the ready incorporation of complex textual responses.  Consequently, these methods may be limited in the types of data they can analyze as well as the numbers of variables that they can practically handle.

Motivated by the problem of identifying potential hierarchical population structure on modern survey data containing a wide range of complex data types, we introduce population-based hierarchical non-negative matrix factorization (PHNMF). PHNMF is a variant of hierarchical non-negative matrix factorization based on feature similarity.  As such, it enables an automatic and interpretable approach for identifying and understanding hierarchical structure in a data matrix constructed from a wide range of data types.  Our numerical experiments in Sections \ref{sec:experiments} and \ref{sec:results} demonstrate that PHNMF can recover latent hierarchical population structure with high accuracy.  Moreover, the recovered subpopulation structure is interpretable and can be useful for improving downstream inference.

In the rest of this paper, we first discuss some related works and highlight our contributions in Section \ref{sec:relatedworks}.  We then set some notation and describe background on non-negative matrix factorization (NMF), its employment in topic modeling, and its hierarchical variant in Section \ref{sec:background}.  We introduce PHNMF and highlight its differences from hierarchical NMF in Section \ref{sec:method}, describe the design of our numerical experiments in Section \ref{sec:experiments}, and present our results in Section \ref{sec:results}.  Finally, we conclude with a brief discussion and some directions for future work in Section \ref{sec:discussion}.

\subsection{Related Work and Contributions}
\label{sec:relatedworks}

Traditional statistical methods for discovering latent subgroups from observed data include latent class analysis (LCA) and latent profile analysis (LPA).  
LCA is a probabilistic method for discovering latent groupings of observed categorical variables and assumes that variables within each group are independent (\!\cite{poLCAreview}, \cite{JSSv042i10}).
LPA is the continuous data version of LCA.  It is an instance of multivariate mixture estimation for continuous variables that likewise assumes that variables within each group are independent (\!\cite{LCAlpa}, \cite{oberski}). One can therefore employ these methods to cluster members of a population by assigning members to the most probable latent class.  While the models obtained from LCA and LPA depict a flat clustering structure, one can also employ the methods recursively within groups to obtain a hierarchical clustering tree.

PHNMF makes several improvements over these methods.  First, PHNMF makes no distributional assumptions on the observed data and does not require independence on subsets of the variables.  Nonetheless, our numerical experiments in Section \ref{sec:experiments} demonstrate that PHNMF can recover subpopulation structure with high accuracy.  
Therefore, PHNMF may better accommodate modern survey data, which can include complex data with non-trivial correlations.

A second advantage of PHNMF is improved interpretability of the resulting clustering structure.  While LCA and LPA return the probabilities that the members belong to particular slatent classes, they treat all variables within a class as equally important.  By contrast, PHNMF returns the relative importance of each variable to a member's responses.  This enables understanding of how the variables contribute to each member's clustering assignments.

A third advantage of PHNMF is that it is designed specifically for discovering hierarchical structure in complex data.  Our numerical results in Section \ref{sec:results} illustrate how LCA and LPA may struggle to identify meaningful structure in complex data with a large number of variables.  By contrast, PHNMF readily discovers interpretable hierarchical population structure that aligns with existing literature.

\section{Background}
\label{sec:background}

We first set some notation for the rest of this paper.  We then present a brief overview of non-negative matrix factorization, its applications to topic modeling, and hierarchical non-negative matrix factorization.


We assume that all vectors are column vectors.  
Given a matrix $\mathbf{X}$, we denote its $(i,j)^{th}$ entry by $\mathbf{X}_{i,j}$, its $i^{th}$ row by $\mathbf{x}_{i}^{T}$, and its Frobenius norm by $\|\mathbf{X}\|_{F}$.  We employ the notation $\mathbf{X} \ge 0$ to indicate that it is non-negative so that all its entries are greater than or equal to $0$.

\subsection{Non-Negative Matrix Factorization (NMF)}

Non-negative matrix factorization (NMF) \cite{lee1999learning} is a linear dimension reduction technique for interpretable analysis of non-negative data. Given a non-negative matrix $\mathbf{X} \in \mathbb{R}^{n \times m}$ and a rank $k \leq \min(n, m)$, NMF
finds $\mathbf{W}\in \mathbb{R}^{n \times k}$ and $\mathbf{H}\in \mathbb{R}^{k \times m}$ that minimize
\begin{equation*} \label{eqn_1}
(\mathbf{W},\mathbf{H}) = \underset{\mathbf{W} \geq 0, \mathbf{H} \geq 0}{\arg\min}\, \frac{1}{2} \left\|\mathbf{X} - \mathbf{WH} \right\|^2_F.
\end{equation*}

Therefore, NMF approximates $\mathbf{X}$ as the product of two lower-dimensional factor matrices, $\mathbf{W}$ and $\mathbf{H}$.
NMF employs the following multiplicative update (MU) rules from \cite{lee1999learning} and \cite{leealgorithms} to compute the factor matrices
\begin{align}
&\mathbf{W}_{i,j} \leftarrow \mathbf{W}_{i,j}\frac{(\mathbf{X}(\mathbf{H})^T)_{i,j}}{(\mathbf{W}\mathbf{H}(\mathbf{H})^T)_{i,j}},\label{eqn_3} \quad \text{and}\\
&\mathbf{H}_{i,j} \leftarrow \mathbf{H}_{i,j}\frac{((\mathbf{W})^T\mathbf{X})_{i,j}}{((\mathbf{W})^T\mathbf{W}\mathbf{H})_{i,j}} \label{eqn_4}.
\end{align}

Alg.~\ref{alg:nmf} presents pseudocode for computing an NMF factorization.  
Compared with other matrix factorization techniques that can produce factor matrices containing negative components, the non-negativity constraint of NMF enables straight-forward interpretation.  
Due to its interpretability, NMF is popular in a wide variety of applications including image processing \cite{CHEN201940} and hyperspectral unmixing \cite{hyperspectral}. 

\begin{algorithm}[ht] 
	\caption{\label{alg:nmf} NMF}
	{\bf Input:} Data matrix $\mathbf{X} \geq 0$, $k \leq \min(n, m)$\\
    \begin{algorithmic}[1]
            \STATE initialize $\mathbf{W} \geq 0 \in \mathbb{R}^{n \times k}, \mathbf{H} \geq 0 \in \mathbb{R}^{k \times m}$
            
            \WHILE{not converged}
                \STATE Compute $\mathbf{W}\mathbf{H}$\\
                \STATE Update $\mathbf{H}$ by MU rules in \eqref{eqn_4}\\
                \STATE Update $\mathbf{W}$ by MU rules in \eqref{eqn_3}
            \ENDWHILE
            \RETURN $\mathbf{W} \geq 0 \in \mathbb{R}^{n \times k},\, \mathbf{H} \geq 0 \in \mathbb{R}^{k \times m}$
	\end{algorithmic}
\end{algorithm}

\subsection{NMF for Topic Modeling}
\label{topicmodelnmf}
Let $\mathbf{X}$ represent a corpus of documents with the documents on the rows and the words in the documents on the columns.  Then NMF recovers topics within the corpus and classifies the topical composition of each document \cite{gillis_nmf}. The resulting NMF factorization produces $\mathbf{W}$ and $\mathbf{H}$ such that $\mathbf{W}_{ij}$ represents the weight of the $j^{th}$ topic in the $i^{th}$ document, and $\mathbf{H}_{ij}$ represents the weight of the $j^{th}$ word in the $i^{th}$ topic. This factorization enables simultaneous identification of latent topics and understanding of the topical composition of each document.

\subsection{Hierarchical NMF (HNMF)}

Hierarchical NMF (HNMF) applies NMF recursively to reveal latent hierarchical topic structure in the data.
Therefore, HNMF can offer a richer understanding than classical NMF, which assumes a flat latent structure.  

There are multiple approaches to performing HNMF.  We consider top-down HNMF, which first discovers high-granularity topics and then recursively discovers subtopics within broader ones \cite{fastrank2NMF}.
It does this by recursively applying Alg.~\ref{alg:nmf}. Concretely, let $k^{(0)}$ denote the rank parameter at the first iteration.  Then top-down HNMF first factorizes $\mathbf{X}$ as
\begin{equation*} \label{topdown}
\mathbf{X} \approx \mathbf{W}_0\mathbf{H}_0,
\end{equation*}
where $\mathbf{H}_0 \in\mathbb{R}^{k^{(0)} \times m}$ describes the composition of the broadest supertopics and $\mathbf{W}_0 \in\mathbb{R}^{n \times k^{(0)}}$ encodes the corresponding representations of each document. At the $i^{th}$ iteration, top-down HNMF splits documents into submatrices $\mathbf{X}^{(i)}_{1}, \mathbf{X}^{(i)}_{2},\dots, \mathbf{X}^{(i)}_{k^{(i)}}$, where $\mathbf{X}^{(i)}_{j}$ for $1 \le j \le k^{(i)}$ contains only the documents that discuss the $j^{th}$ topic as determined by the coefficients in $\mathbf{H}_{i}$ and a minimum discussion threshold $\alpha$. Top-down HNMF then applies NMF to each of these submatrices to discover the subtopics that form the $j^{th}$ supertopic.
The algorithm continues until subtopics no longer contain some minimum number of documents $t$.  Alg.~\ref{alg:topdown-hnmf} presents pseudocode for these procedures. 
\begin{algorithm}[ht]
	\caption{\label{alg:topdown-hnmf} Top-down HNMF}
	{\bf Input:} Data matrix $\mathbf{X} \in \mathbf{R}^{n \times m}$ with $\mathbf{X} \geq 0$,  
	minimum topic discussion threshold $\alpha$, minimum number of documents in a topic $t$

    \begin{algorithmic}[1]
	    \STATE Initialize $i\leftarrow 0$
        \STATE Determine $k^{(i)}$, number of latent topics in $\mathbf{X}$  \label{alg:hnmf1}
		\STATE Compute $(\mathbf{W}_{i},\,\mathbf{H}_{i}) = \text{NMF}(\mathbf{X}, k^{(i)})$\\
        \FOR{each document $1 \le l \le n$ in $\mathbf{X}$}
            \STATE Assign document $l$ to topic submatrix
            $\mathbf{X}^{(i)}_j$ if $(\mathbf{W}_{i})_{l,j} > \alpha$
        \ENDFOR
            \FOR{each topic submatrix $\mathbf{X}_{j}^{(i)}$}       
            \WHILE{number of documents in topic $j > m$}
                \STATE Update $i \leftarrow i+1$
                \STATE Determine $k^{(i)}$, number of latent topics in $\mathbf{X}_{j}^{(i-1)}$ \label{alg:hnmf2}\\
                \STATE Compute $(\mathbf{W}_{i},\,\mathbf{H}_{i}) = \text{NMF}(\mathbf{X}_{j}^{(i-1)}, k^{(i)})$ \\
                \STATE Assign documents in $\mathbf{X}_{j}^{(i-1)}$ to topic submatrices based on whether corresponding coefficients in $\mathbf{W}_{i} > \alpha$
            \ENDWHILE
        \ENDFOR
	\end{algorithmic}
\end{algorithm}

Lines~\ref{alg:hnmf1} and \ref{alg:hnmf2} in Alg.~\ref{alg:topdown-hnmf} involve determining an input rank $k$ in Alg.~\ref{alg:nmf}.  While Alg.~\ref{alg:nmf} accommodates any method for choosing the rank, we highlight the method in \cite[Algorithm 2]{grotheerhnmf}. This method computes NMF factorizations with each of  $k\in[2,9)$ topics and produces a score between 0 and 1 for each $k$ based on the cohesiveness of the resulting topics.  Higher scores indicate better model fit.

\section{Population-Based HNMF (PHNMF)}
\label{sec:method}

We now present population-based hierarchical NMF (PHNMF), a variation on HNMF that automatically discovers latent population hierarchical structure based on feature similarity, where similarity is based on the pairwise cosine similarity as defined in \cite[Section 5.1]{grotheerhnmf}.  Suppose that $\mathbf{X}$ is a data matrix with individuals on the rows and features derived from responses to survey questions on the columns.   
At each iteration, PHNMF divides the population into disjoint subpopulations by whether a respondent's coefficients for a subpopulation exceeds a threshold quantity $\alpha$. The coefficients in the $\mathbf{W}$ and $\mathbf{H}$ factor matrices at each iteration determine population assignments and relative importance of the discovered features, respectively.  Rather than terminating when the number of observations in a subpopulation falls below a threshold quantity, however, PHNMF terminates when the feature similarity falls below a threshold amount. 

At a high level, feature similarity measures the pairwise cosine similarity between the rows of the $\mathbf{H}$ factor matrices obtained from multiple NMF runs on the same subpopulation with a pre-determined optimal rank and different initializations.  To compute feature similarity, we employ a modified version of \cite[Algorithm 2]{grotheerhnmf}.  Rather than iterating over multiple ranks $k$, we employ only the pre-determined optimal rank. Additionally, rather than returning the maximum of the median value of the least similarity seeds, we return the minimum of the minimum of the least similarity seeds.

Alg.~\ref{alg:ours} presents pseudocode for PHNMF.  PHNMF differs from HNMF in two key ways. First, PHNMF performs ``hard" splits so that each level produces disjoint subpopulations.  This ensures that each respondent belongs to at most one subpopulation at each level.  Second, PHNMF employs feature similarity as the stopping criteria rather than terminating when each subpopulation acquires a minimum number of respondents.  This feature similarity criteria simultaneously avoids arbitrary termination and produces more interpretable subpopulation discovery. 

\begin{algorithm}[ht]
	\caption{\label{alg:ours} Population-Based HNMF (PHNMF)} 
	{\bf Input:} Data matrix $\mathbf{X} \in \mathbf{R}^{n \times m}$ with $\mathbf{X} \geq 0$, 
	minimum subpopulation threshold $\alpha$, minimum feature similarity threshold $\beta$
	
    \begin{algorithmic}[1]
	    \STATE Initialize $i\leftarrow 0$
        \STATE Determine $k^{(i)}$, number of latent subpopulations in $\mathbf{X}$  
		\STATE Compute $(\mathbf{W}_{i},\,\mathbf{H}_{i}) = \text{NMF}(\mathbf{X}, k^{(i)})$\\
        \FOR{each respondent $1 \le l \le n$ in $\mathbf{X}$}
            \STATE Compute $j = \max_{1 \le f \le m} (\mathbf{W}_{i})_{l,f}$
            \STATE Assign respondent $l$ to population submatrix
            $\mathbf{X}^{(i)}_j$ if $(\mathbf{W}_{i})_{l,j} > \alpha$
        \ENDFOR
        \FOR{each population submatrix $\mathbf{X}_{j}^{(i)}$} 
            \STATE Compute feature similarity for $\mathbf{X}_{j}^{(i)}$\\
            \WHILE{feature similarity $> \beta$}
                \STATE Update $i \leftarrow i+1$
                \STATE Determine $k^{(i)}$, number of latent subpopulations in $\mathbf{X}_{j}^{(i-1)}$ 
                \\
                \STATE Compute $(\mathbf{W}_{i},\,\mathbf{H}_{i}) = \text{NMF}(\mathbf{X}_{j}^{(i-1)}, k^{(i)})$ \\
                \STATE Assign respondents in $\mathbf{X}_{j}^{(i-1)}$ to population submatrices based on whether corresponding coefficients in $\mathbf{W}_{i} > \alpha$
            \ENDWHILE
        \ENDFOR
	\end{algorithmic}
\end{algorithm}

\section{Experimental Setup}
\label{sec:experiments}

We demonstrate PHNMF on both synthetic and real data.  Our synthetic data mimic responses from distinct subpopulations to both multiple choice and open-form questions.  We also employ real data from two survey datasets. 

\subsection{Synthetic Data}
\label{sec:synthdata}

We construct two synthetic datasets with hierarchical population structure containing continuous and categorical variables, respectively.  These data mimic the numerical encoding of responses to multiple choice and open-form survey questions.  Each dataset consists of a pair of synthetic explanatory variables and response vector.  We describe our procedures for constructing each below.

\subsubsection{Synthetic Continuous Data}

\begin{figure*}[t]
    \centering
    \includegraphics[width = \textwidth]{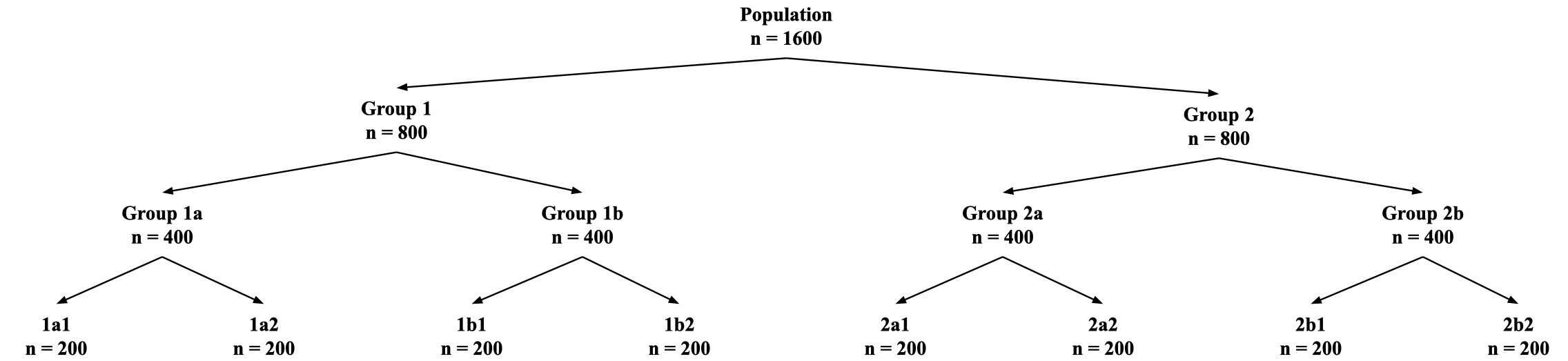}
    \caption{\label{fig:conttree}\emph{Hierarchical population structure of synthetic datasets.} }
\end{figure*}

We construct synthetic explanatory variables $\mathbf{X}^{\text{I}} \in \mathbb{R}^{1600 \times 120}$ containing $1600$ observations on $120$ continuous features with hierarchical population structure as described in Fig.~\ref{fig:conttree}.  In the context of topic modeling for survey data, these data mimic a population containing eight groups, each containing $200$ respondents.  Each respondent discusses some combination of four topics and each topic consists of $30$ words.  While we describe these data in the context of topic modeling, our approach also applies to continuous variables in general. 

We construct $\mathbf{X}^{\text{I}}$ as the product of two rank-$4$ matrices with
\begin{align*}
    \mathbf{X}^{\text{I}} = \mathbf{W}^{\text{I}}\mathbf{H}^{\text{I}},
\end{align*}
where $\mathbf{W}^{\text{I}} \in \mathbf{R}^{1600 \times 4}$ 
is a person-topic matrix and $\mathbf{H}^{\text{I}} \in \mathbf{R}^{4 \times 120}$ 
is a topic-word matrix.  Therefore, $\mathbf{W}_{i,j}^{\text{I}}$ reflects the amount that the $j^{th}$ topic is discussed by the $i^{th}$ person, and $\mathbf{H}_{i,j}^{\text{I}}$ reflects the importance of the $j^{th}$ word in the $i^{th}$  topic. 

We construct $\mathbf{W}^{\text{I}}$ according to the hierarchical structure depicted in Fig.~\ref{fig:conttree}. At the first split, Groups 1 and 2 discuss words in Topic 1 according to zero-truncated $\mathcal{N}(64, 9)$ and $\mathcal{N}(3, 9)$ distributions, respectively.  At the second split, Groups 1a and 2a and Groups 1b and 2b discuss words in Topic 2 according to zero-truncated $\mathcal{N}(45, 9)$ and $\mathcal{N}(3,9)$ distributions, respectively.  At the third split, Groups 1a1, 2a1, 1b1, and 2b1 and Groups 1a2, 2a2, 1b2, and 2b2 discuss words in Topic 3 according to zero-truncated $\mathcal{N}(3,9)$ and $\mathcal{N}(50,9)$ distributions, respectively. Finally, the entire population discusses words in Topic 4 according to a zero-truncated $\mathcal{N}(50,9)$ distribution.  

\begin{figure}[t]
    \centering
    \includegraphics[width = .45\textwidth]{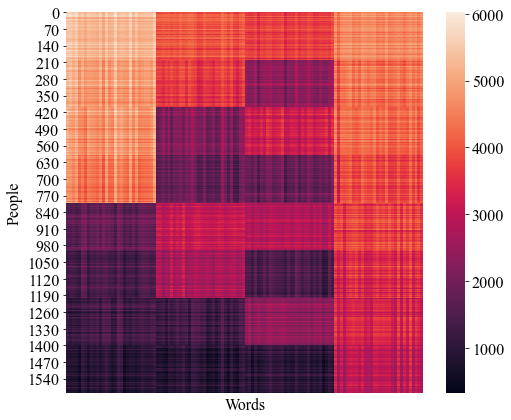}
    \caption{\label{fig:Xmatrix} \emph{Heatmap depicting  structure of synthetic continuous data matrix $\mathbf{X}^{\text{I}}$.} 
    }
\end{figure}

We construct $\mathbf{H}^{\text{I}}$ as follows. Suppose that each of the four topics consists of 30 words. For simplicity, we assume that the words in each topic appear together in the columns of $\mathbf{H}^{\text{I}}$.   
For each topic, we construct 30 vectors of length four by sampling from a multinomial distribution so that the words in each topic are four times more likely to appear in that topic than in the others.
This simulates a more disjoint vocabulary; words are less likely to appear in multiple topics simultaneously.  
Fig.~\ref{fig:Xmatrix} depicts the structure in the resulting synthetic explanatory variables matrix $\mathbf{X}^{\text{I}}$.

We construct a synthetic coefficient vector $\boldsymbol{\theta}_{g} \in \mathbb{R}^{4}$ with $1 \le g \le 8$ for the subgroups. We obtain a synthetic response $\mathbf{y}^{\text{I}} \in \mathbb{R}^{1600}$ by multiplying the corresponding rows of $\mathbf{W}^{\text{I}}$ with $\boldsymbol{\theta}_{g}$ for $1 \le g \le 8$. This simulates a response that is determined by different combinations of the topics for each subgroup.

\subsubsection{Synthetic Categorical Data}

We construct synthetic explanatory variables $\mathbf{X}^{\text{II}} \in \mathbb{R}^{1600 \times 120}$ containing $1600$ observations on $120$ dichotomous categorical features with hierarchical population structure as described in Fig.~\ref{fig:conttree}.  We construct $\mathbf{X}^{\text{II}}$ according to
\begin{align*}
    \mathbf{X}^{\text{II}} = \text{threshold}(\mathbf{W}^{\text{I}}\mathbf{H}^{\text{II}}),
\end{align*}
where we employ the same $\mathbf{W}^{\text{I}}$ and construct $\mathbf{H}^{\text{II}}$ similarly to $\mathbf{H}^{\text{I}}$.  We threshold the entries in the product $\mathbf{W}^{\text{I}}\mathbf{H}^{\text{II}}$ so that $\mathbf{X}^{\text{II}}$ contains only 1's and 0's.  

We highlight some differences in the procedures for constructing $\mathbf{H}^{\text{II}}$.  Rather than employing 30 words in each topic as in $\mathbf{H}^{\text{I}}$, we vary the number of words in each topic to ensure ordered topical importance.  To see this, notice that a topic's relative importance is determined by the magnitudes of its entries and its size, or the number of words it contains.  For example, equal numbers of entries containing 1's and 0's and equal topic sizes produces equally important topics.  
Therefore, we vary the number of words in each topic to induce an importance hierarchy.
Topic 1 consists of 65 words, Topic 2 consists of 30 words, Topic 3 consists of 20 words, and Topic 4 consists of 5 words.  We again assume that words appear together by topic.   

\begin{figure}[th]
    \centering
    \includegraphics[width = 0.45\textwidth]{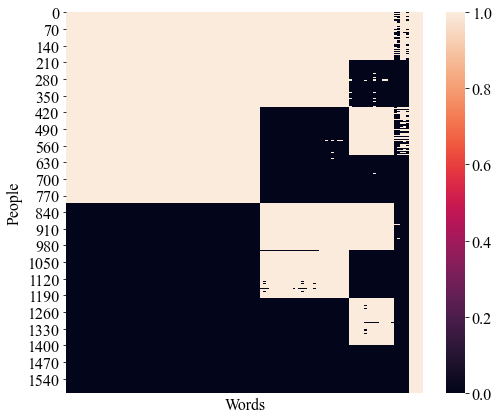}
    \caption{\label{fig:Xcat} \emph{Heatmap depicting structure of synthetic categorical data matrix $\mathbf{X}^{\text{II}}$.} 
    }
\end{figure}

To ensure that $\mathbf{X}^{\text{II}}$ contains only dichotomous categorical variables, we threshold the entries in the product $\mathbf{W}^{\text{I}}\mathbf{H}^{\text{II}}$ as follows.  Suppose that the $j^{th}$ word belongs to the $l^{th}$ topic and let $\text{med}_{l}$ denote the median value of the words in $\text{Topic} \, l$.  Then we obtain $\mathbf{X}^{\text{II}}_{i,j}$ with
\begin{align*}
    \mathbf{X}^{\text{II}}_{i,j} = 
        \begin{cases}
            1 \quad \text{if $(\mathbf{W}^{\text{I}}\mathbf{H}^{\text{II}})_{i,j} >$ med$_{l}$}, \quad \text{and}\\
            0 \quad \text{otherwise}.
        \end{cases}
\end{align*}
Fig.~\ref{fig:Xcat} depicts the structure in $\mathbf{X}^{\text{II}}$. We employ the same response variable $\mathbf{y}^{\text{I}}$ 
since the hierarchical population structure determined by $\mathbf{W}^{\text{I}}$ remains unchanged.

\subsection{Real Data}

\subsubsection{City of Austin Satisfaction Survey}

We employ real data containing survey responses from residents of Austin, Texas on their satisfaction with various city services \cite{aus}. 
Our dataset consists of the $3{,}723$ completed surveys obtained since 2018 that do not contain any empty columns. 

Approximately 200 columns correspond to multiple choice questions such as the following.
\begin{enumerate}
   \item Overall quality of public safety services (i.e. police, fire and ambulance) \emph{(Choices: Very Dissatisfied, Dissatisfied, Neutral, Satisfied, Very Satisfied, Don't Know)}
   \item I feel safe in my neighborhood during the day \emph{(Choices: Strongly Disagree, Disagree, Neutral, Agree, Strongly Agree, Don't Know)}
   \item Have you had contact with the City of Austin Municipal Court? \emph{(Choices: Yes, No, Don't Know)}
   \item Which two items listed in Quality of Life do you think are most important for the City to provide?  \emph{(Choices: `Feel Welcome', `Overall quality of life', `Place to Live', `Place to Work', `Raise Children', `Retire', `None')}
\end{enumerate}
In addition to the multiple choice questions, we include eight columns corresponding to open-form response questions on respondent satisfaction levels.   We exclude open-form questions that closely resemble the multiple choice questions.

We consolidate multiple choice responses on satisfaction to positive, neutral, and negative responses. We construct indicator variables for positive and negative responses and retain these as input variables. For each open-form question, we employ \verb|tfidfvectorizer| from \verb|scikit-learn| in Python \cite{scikit-learn} to convert text responses to a respondent-word matrix and perform NMF to identify the primary response topics. For each topic, we construct a new variable whose entries are the corresponding NMF coefficients indicating how much each respondent discusses each topic, and set the largest coefficient to $1$. The resulting dataset contains $3{,}723$ observations on $250$ continuous and categorical variables.  We exclude demographic variables in performing PHNMF and employ them only to ascertain the meaningfulness of the PHNMF subgroups.

\subsubsection{Facebook Climate Change Survey}

We employ real data containing multiple choice questions from a Facebook survey measuring public opinion on climate change from $2{,}432$ respondents \cite{zhangfacebookquota}.  We emphasize that these data and results do not reflect our personal views on climate change. We apply PHNMF to the following seven multiple choice questions.
\begin{enumerate}
   \item Do you think that global warming is happening? \emph{(Choices: Yes, No, Don't know)}
   \item Assuming global warming is happening, do you think it is mostly… \emph{(Choices: Caused by human activities, Caused by natural environmental changes, Other)}
   \item Which comes closest to your own view? \emph{(Choices: Most scientists think global warming is not happening, There is much disagreement among scientists about whether or not global warming is happening, Most scientists think global warming is happening, Don’t know)}
    \item How much do you think global warming will harm you personally? \emph{(Choices: A great deal, A moderate amount, Only a little, Not at all, Don't know)}
   \item How much do you think global warming will harm future generations of people? \emph{(Choices: A great deal, A moderate amount, Only a little, Not at all, Don't know)}
   \item How much do you think global warming will harm plant and animal species? \emph{(Choices: A great deal, A moderate amount, Only a little, Not at all, Don't know)}
   \item How much do you support or oppose funding more research into renewable energy sources, such as solar and wind power. \emph{(Choices: Strongly support, Somewhat support, Somewhat oppose, Strongly oppose)}
 \end{enumerate}

For each question, we construct indicator variables for each multiple choice response.  The resulting dataset contains $2{,}427$ observations on $35$ categorical variables.  Again, we exclude demographic variables when performing PHNMF and employ them only to investigate the meaningfulness of the PHNMF subgroups.   

\subsection{Experimental Setup}

We perform two sets of experiments to demonstrate the performance of our proposed method.  1) First, we perform experiments to test PHNMF's ability to recover latent hierarchical clustering structure on both continuous and categorical data.  2) Second, we perform experiments to illustrate the potential usefulness of the recovered clustering structure for improved inference with linear and ridge regression as appropriate.

We perform 1,000 replicates with PHNMF for the continuous and categorical synthetic data described in Sec.~\ref{sec:synthdata}. For each replicate, we construct a new pair of synthetic explanatory and response variables and record the classification accuracy.

We compare our results to those obtained from applying LPA and LCA for continuous and categorical data, respectively.  We apply LPA to continuous data with both the \verb|mclust| and \verb|tidyLPA| packages for R (\!\cite{fraley1998mclust}, \cite{rosenberg2019tidylpa}).  We apply LCA to categorical data with the \verb|poLCA| package for R \cite{poLCA}.  
While Algorithm \ref{alg:ours} and our synthetic data construction employ Python, the packages for LCA and LPA are in R.  LCA and LPA also require substantial time to determine an optimal number of classes for each dataset.  Therefore, we perform 300 replicates with LCA and LPA.
We run 10 replicates of LPA for each of $1 \le k \le 10$ classes to account for randomness in the initialization and select the model with lowest BIC \cite{neath2012bayesian}.

To illustrate the potential usefulness of the recovered clustering structure for improved inference with synthetic data, we perform regression with the explanatory and response variables for each subpopulation discovered by PHNMF and the population as a whole.  We record the cosine similarity between the resulting coefficient vectors and both the known coefficient vectors in Sec.~\ref{sec:synthdata} and the coefficient vector we obtain from regression with the entire population.  We construct the continuous synthetic data so that all the subgroups have full column rank in the explanatory variables to avoid issues with identifiability.  Therefore, we employ linear regression for the continuous data.  Since the categorical synthetic data may lack full column rank in the subgroups, we instead employ ridge regression.

Since the real datasets lack known coefficient vectors, we select the Facebook Climate Change Survey to illustrate the potential usefulness of the recovered clustering for downstream inference.   We employ an ordinal variable containing a self-identified political ideology score ranging from 1 to 5, where 1 is ``very liberal" and 5 is ``very conservative". 
Since some subpopulations discovered by PHNMF may lack full column rank in the explanatory variables,  we perform ordinal ridge regression \cite{ordinalridge} with the \verb|mord| 
package in Python.  We employ three iterations of 5-fold cross-validation over $\lambda = [0.0001, 0.001, 0.01, 0.01, 1]$ to select the $\lambda$ that minimizes the mean squared error for the ridge penalty.

\section{Results}
\label{sec:results}

We present the results of our numerical experiments on synthetic and real datasets, respectively.

\subsection{Synthetic Data Results}

\begin{figure}[htbp]
    \begin{center}
    \includegraphics[width = 0.45\textwidth]{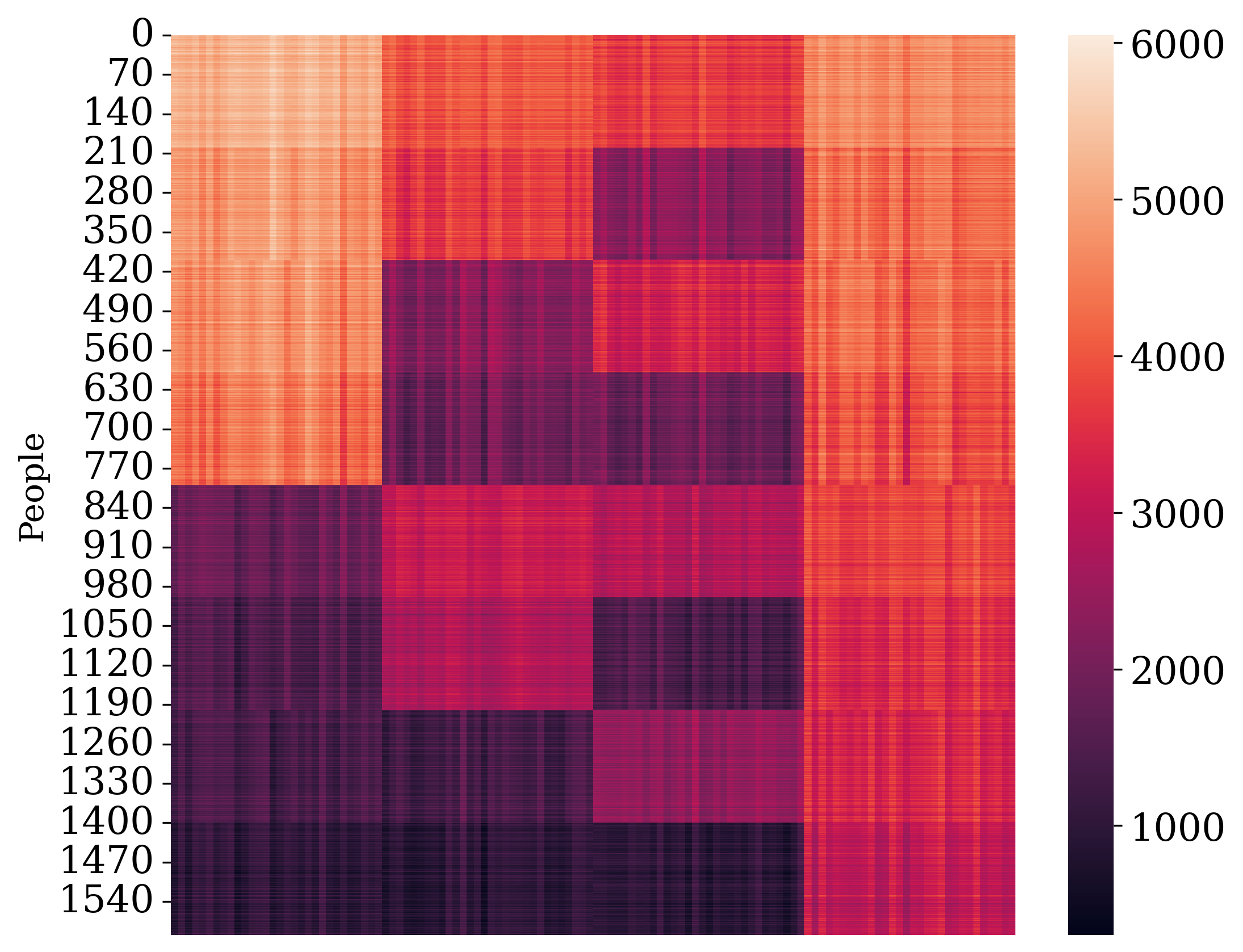}
    \end{center}
    \caption{\label{fig:recoverX} \emph{Heatmap depicting structure of recovered and sorted synthetic continuous data matrix $\mathbf{X}^{\text{I}}$.}
    }
\end{figure}

\begin{figure}[htbp]
    \begin{center}
    \includegraphics[width = 0.45\textwidth]{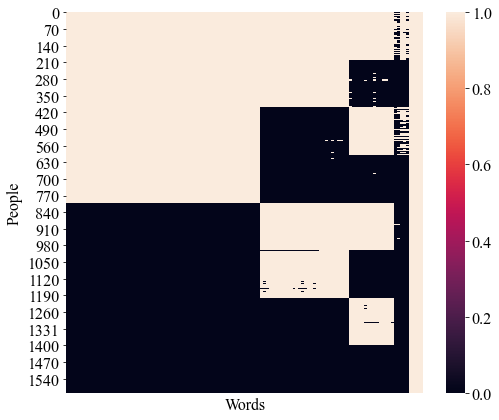}
    \end{center}
    \caption{\label{fig:recoverXcat} \emph{Heatmap depicting structure of recovered and sorted synthetic categorical data matrix $\mathbf{X}^{\text{II}}$.}
    }
\end{figure}

Fig.~\ref{fig:recoverX} and Fig.~\ref{fig:recoverXcat} depict data matrices recovered from a single replicate of the synthetic continuous and categorical data with PHNMF, respectively.  They are qualitatively similar to Fig.~\ref{fig:Xmatrix} and Fig.~\ref{fig:Xcat}.

Tab.~\ref{tab:synthresults} depicts results of experiments testing PHNMF's ability to recover latent hierarchical population structure on both continuous and categorical data. Notice that different runs of our algorithm may return subgroups in a different ordering since a  factorization $\mathbf{WH} = \mathbf{WP}^{T}\mathbf{PH}$, where $\mathbf{P}$ is a permutation matrix. Therefore, we identify the eight subgroups by defining each group by the most frequently observed label for observations assigned to the group.

Tab.~\ref{tab:synthresults} shows that PHNMF recovers the latent hierarchical structure in both datasets with high accuracy.  LCA performs nearly as well as PHNMF on the categorical data but does not handle continuous data.  We attempt LPA with two R packages but neither produces functional results; \verb|tidyLPA| assigns all observations to a single group while \verb|mclust| does not run to completion.  Since the packages can return results on subsets of the columns from the synthetic continuous datasets, it appears that the full datasets contain more variables than the LPA algorithms can practically handle.

\begin{table}[htbp]
\caption{Clustering accuracy  (mean and standard error) over $1{,}000$ replicates of PHNMF and $300$ replicates of LCA and LPA on synthetic datasets. We do not report LPA results since \texttt{tidyLPA} assigns all observations to a single group and \texttt{mclust} does not run to completion.}
\begin{center}
\begin{tabular}{lcccc}
\hline
\multicolumn{1}{c}{}                & \multicolumn{2}{c}{\textbf{Continuous Data}}                                        & \multicolumn{2}{c}{\textbf{Categorical Data}}           
\\
\hline
\multicolumn{1}{c}{\textbf{Method}} & \multicolumn{1}{c}{\textbf{Accuracy}} & \multicolumn{1}{c}{\textbf{Std. Err.}} & \multicolumn{1}{c}{\textbf{Accuracy}} & \multicolumn{1}{c}{\textbf{Std. Err.}} \\
\hline \\
PHNMF & 98.5   & 0.04 & 99.97 & 0.0004\\
LCA & -- & -- & 96.43 & 0.059\\
LPA & -- & -- & -- & --  \\
\hline
\end{tabular}
\end{center}
\label{tab:synthresults}
\end{table}

Tab.~\ref{tab:thetaresults} presents results on the potential usefulness of the recovered PHNMF clustering structure for improved downstream inference from a single replicate of the continuous and categorical datasets, respectively.  Since the cosine similarity computes the cosine of the angle between two vectors, it is $1$ when two vectors are perfectly aligned and $0$ when they are orthogonal.  Columns 2 and 4 depict the cosine similarity between the known coefficient vectors and those recovered from regression on the subgroups for the synthetic continuous and categorical datasets, respectively.  Columns 3 and 5 depict the cosine similarity between the coefficient vectors obtained from 
regression on the entire population and those recovered from regression on the subgroups for the synthetic continuous and categorical datasets, respectively.

\begin{table}[htbp]
\caption{Cosine similarity between recovered coefficient vectors from regression on the subgroups compared to: 1) the known subgroup coefficient vectors and 2) the recovered coefficient vectors from regression with the entire population.  Cosine similarity $1$ indicates perfect directional alignment; $0$ indicates orthogonality. 
}
\begin{center}
\begin{tabular}{lllll}
\hline
\multicolumn{1}{c}{}                & \multicolumn{2}{c}{\textbf{Continuous Data}}                                        & \multicolumn{2}{c}{\textbf{Categorical Data}}           
\\
\hline
\multicolumn{1}{c}{\textbf{Group Name}} & \multicolumn{1}{c}{\textbf{Subgroups}} & \multicolumn{1}{c}{\textbf{Population}} & \multicolumn{1}{c}{\textbf{Subgroups}} & \multicolumn{1}{c}{\textbf{Population}} \\

\hline \\
Group 1a1 &0.8793 &0.1020 &0.9320 &0.4773\\
Group 1a2 &0.9730 &0.2866 &0.9157 &0.5234 \\
Group 1b1 &0.9310 &0.7022 &0.9596 &0.5008\\
Group 1b2 &0.9271 &0.0789
 &0.9117 &0.5095 \\
Group 2a1 &0.9546 &0.9229
 &0.9121 &0.3572\\
Group 2a2 &0.9642 &0.9471 &0.9959 &0.3265 \\
Group 2b1 &0.9222 &0.8786 &0.9712 &0.4728 \\
Group 2b2 &0.8023 &0.3716
 &01.000 &0.7988 \\
\hline
\end{tabular}
\end{center}
\label{tab:thetaresults}
\end{table}

Tab.~\ref{tab:thetaresults} shows that coefficient vectors obtained on the PHNMF subgroups align more closely with the known coefficient vectors.  This gives evidence that the subpopulation structure discovered by PHNMF may improve downstream inference.

\subsection{Real Survey Data Results}

\subsubsection{City of Austin Satisfaction Survey}

\begin{figure}[t]
    \centering
    \includegraphics[width = 0.45\textwidth]{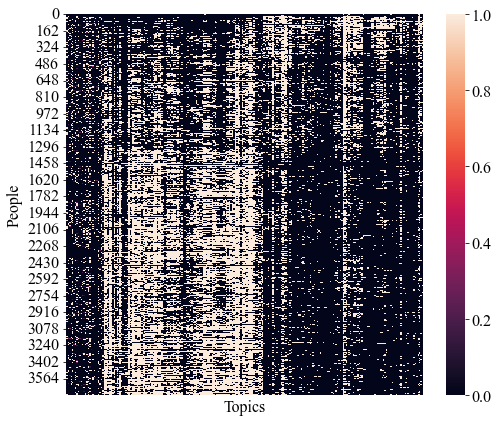}
    \caption{\label{fig:AustinHeatmapResults} \emph{Heatmap depicting structure discovered from PHNMF in City of Austin Satisfaction Survey dataset.}  }
\end{figure}

Fig.~\ref{fig:AustinHeatmapResults} depicts a heatmap of the population structure discovered with PHNMF on the City of Austin Satisfaction Survey. The heatmap suggests non-trivial structure on the columns and the rows.

\begin{figure}[t]
    \centering
    \includegraphics[width =.45 \textwidth]{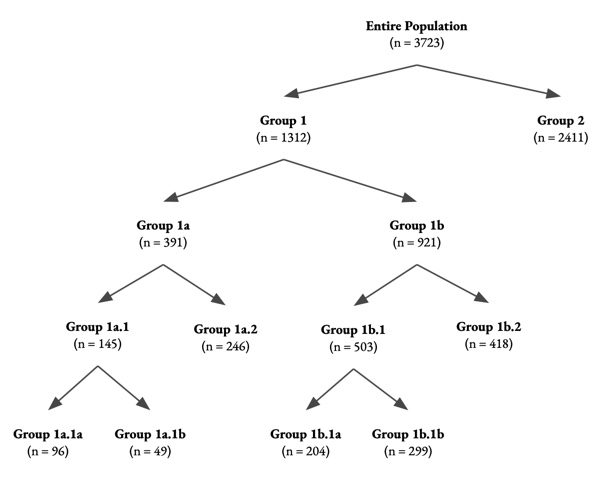}
    \caption{\label{fig:Austintree} \emph{Hierarchical population structure discovered with PHNMF in City of Austin Satisfaction Survey dataset.} }
\end{figure}

Fig.~\ref{fig:Austintree} depicts the hierarchical population structure discovered with PHNMF on the same dataset. The population splits are interpretable and also provide an understanding of differences in survey responses among respondents.   
The first PHNMF split divides the $3{,}723$ survey respondents into two subgroups with $1{,}312$ and $2{,}411$ respondents, respectively. Group 1 primarily responds negatively to multiple choice questions on satisfaction with city services, while Group 2 primarily responds positively to the same questions.

At the second level, PHNMF divides Group 1 into two subgroups with $391$ and $921$ respondents, respectively.  
While Group 1a is distinguished by high levels of dissatisfaction, Group 1b reveals affirmative responses to feeling safe in Austin and satisfaction with the fire department and garbage collection services.
We compare demographic statistics on the PHNMF subgroups and find that compared with the entire population, Group 1a reports lower average income, lower full-time employment rates, and lower percentages of owning their home compared to the entire population.  Meanwhile, Group 1b reports comparable statistics to that of the entire population in these measures.

At the third level, PHNMF divides Group 1a into two groups with $145$ and $246$ respondents, respectively, and Group 1b into two groups with $503$ and $418$ respondents, respectively. Group 1a.1 is distinguished by consistent dissatisfaction.  Meanwhile, Group 1a.2 expresses dissatisfaction with traffic flow and city planning but also expresses satisfaction with Austin’s emergency services, and affirms feeling safe in the city. Group 1b.1 is distinguished by satisfaction with city libraries and Austin as a place to live and feel welcome, and only express dissatisfaction with traffic flow. By contrast, Group 1b.2 has higher rates of dissatisfaction, but affirms feeling safe at home and in their neighborhood, and trusts emergency services. 

At the fourth level, PHNMF splits Group 1a.1 into two groups with $96$ and $51$ respondents, respectively, and Group 1b.1 into two groups with $204$ and $299$ respondents, respectively.  Group 1a.1a and Group 1a.1b consistently respond negatively but Group 1a.1a is more concerned with traffic and city planning.  Meanwhile, Group 1a.1b primarily expresses dissatisfaction with the City of Austin's communications with its residents. By contrast, Group 1b.1a and Group 1b.1b mainly express satisfaction. Group 1b.1a is satisfied with the city library and learning centers, while Group 1b.1b is satisfied with Austin as a place to live and feel welcome, and report feeling safe in the city.

Since $27$ of the $250$ columns correspond to continuous variables, we convert these to categorical variables to employ LCA.  However, LCA struggles to identify the best number of latent classes $k$ in this dataset.  When we enable $k$ to range from $1$ to $15$, LCA reports insufficient degrees of freedom to select an optimal number of classes for $k \ge 13$.  When we limit the number of classes to $k \le 12$, however, LCA selects $k=12$.  Since LCA does not detail the importance of the variables in contributing to each class, the resulting 12 classes on the $250$ columns are difficult to interpret.  Since a small percentage of the variables in this dataset are continuous, we do not employ LPA.

\subsubsection{Facebook Climate Change Survey}

\begin{figure}[th]
    \centering
    \includegraphics[width = 0.45\textwidth]{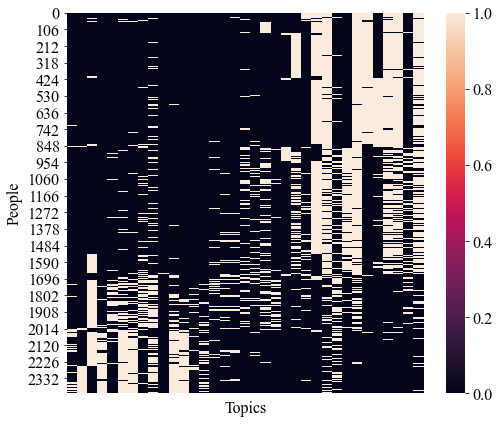}
    \caption{\label{fig:FBHeatmapResults} \emph{Heatmap depicting structure discovered with PHNMF in Facebook Climate Survey dataset.} }
\end{figure}

Fig.~\ref{fig:FBHeatmapResults} depicts a heatmap of the population structure discovered with PHNMF on the Facebook Climate Change Survey.  The heatmap depicts non-trivial structure on the columns and the rows.

\begin{figure*}[t]
    \centering
    \includegraphics[width = \textwidth]{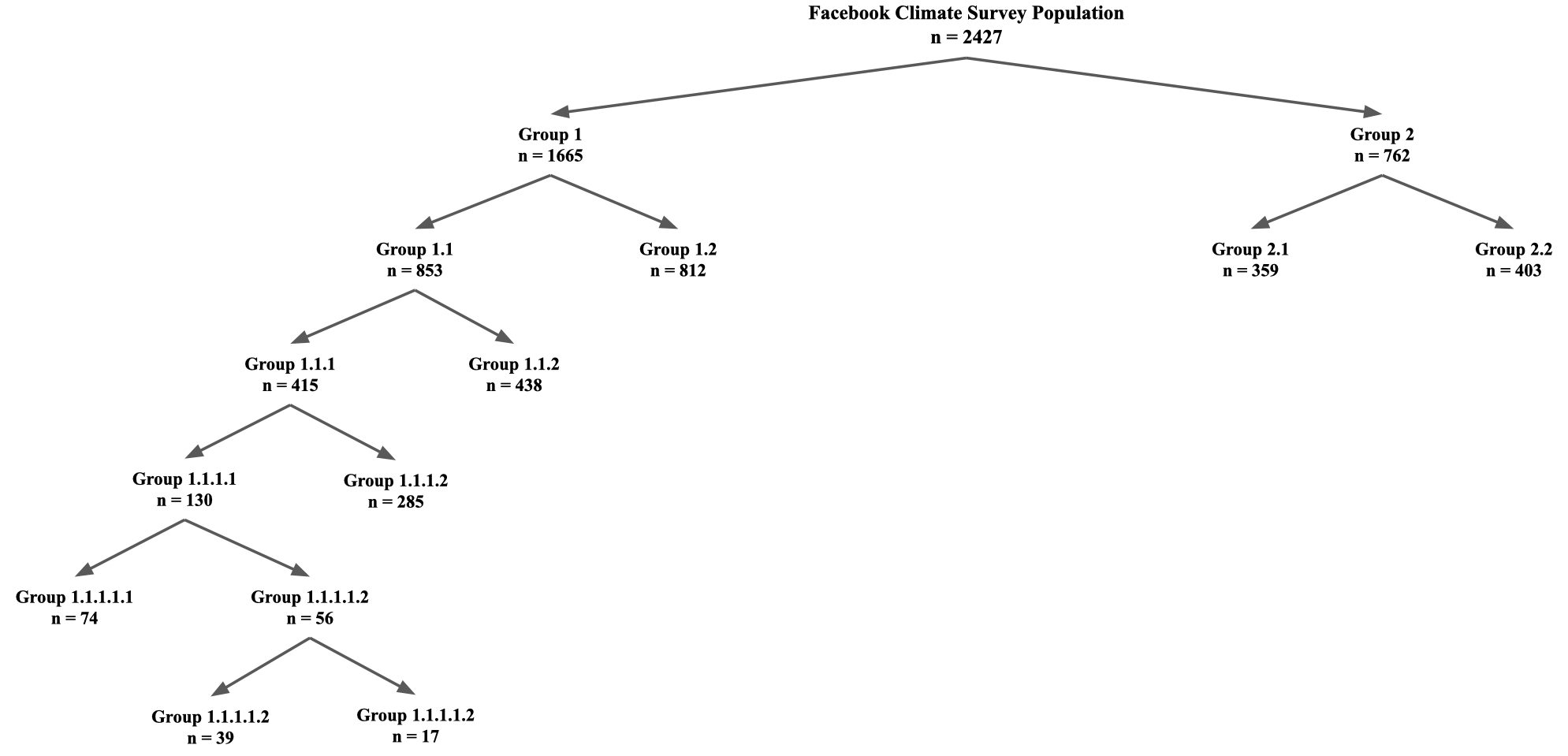}
    \caption{\label{fig:FBTree} \emph{Hierarchical population structure discovered with PHNMF in Facebook Climate Survey dataset.} }
\end{figure*}

Fig.~\ref{fig:FBTree} depicts the hierarchical population structure discovered with PHNMF on the same dataset.  The population splits are interpretable and also align with existing findings on climate change.  At the first split, PHNMF divides the $2{,}427$ survey respondents into two subgroups with $1{,}665$ and $762$ respondents, respectively. Group 1 primarily answers affirmatively to whether global warming exists and expresses strong support for funding renewable energy.  Meanwhile, Group 2 primarily indicates that global warming is caused by natural environmental changes and is unworried about these changes. 
We compare demographic statistics on the PHNMF subgroups and find that Group 1 is approximately 60\% female while Group 2 is approximately 60\% male. This aligns with findings that women are more likely to affirm and express greater concern over climate change (\cite{genderclimate}, \cite{MCCRIGHT20111163}). Additionally, about 60\% of Group 1 is under the age of 45 while about 60\% of Group 2 is over the age of 45. This aligns with findings that younger generations are more likely to support a shift to renewable energy \cite{funk_2021}.   
Finally, 44\% of Group 1 identifies as Democratic and 7\% as Republican 
whereas 4\% of Group 2 identifies as Democratic  40\% as Republican. 
This aligns with findings that Democrats tend to express greater concern over climate change \cite{kennedy_johnson_2020}.

At the second split, PHNMF divides Group 1 into subgroups Group 1.1 and Group 1.2 with $853$ and $812$ respondents, respectively. The distinguishing factor between these subgroups is their level of concern for global warming. Group 1.1 expresses a great deal of worry while Group 1.2 expresses being only somewhat worried.  
The percentage of women in Group 1.1, Group 1.2, and the population at large is 64\%, 56\%, 50\%, respectively.  This aligns with findings of greater levels of climate change concern among women \cite{ballew_marlon_leiserowitz_maibach_2018}. 
Additionally, Group 1.1 consists of 51\% Democrats and 4\% Republicans while Group 1.2 consists of 37\% Democrats and 10\% Republicans. This aligns with findings that Democrats express greater concern over climate change \cite{kennedy_johnson_2020}.
At the second split, PHNMF also divides Group 2 into Groups 2.1 and 2.2 with $359$ and $403$ respondents, respectively. 
Group 2.1 acknowledges the existence of global warming to some extent while Group 2.2 denies global warming altogether.

The third split divides Group 1.1 into Group 1.1.1 and Group 1.1.2 with $415$ and $438$ respondents, respectively. Group 1.1.1 upholds that global warming is occurring and expresses concern, but only expresses moderate concern for the harm that global warming will cause them. Group 1.1.2 identifies that global warming will harm themselves, the planet, and future generations. Members of Group 1.1.2 are about 50\% more likely to be under the age of 18 than those of Group 1.1.1.  This aligns with findings that younger individuals are more likely to express concern that climate change will harm them \cite{bell_poushter_fagan_huang_2022}.  Additional splits are likewise interpretable and appear to align similarly with existing findings on climate change views.

The corresponding LCA analysis finds $14$ classes within the $35$ variables on the first split.  The large number of classes is difficult to interpret.  When we enforce only two classes in the first LCA split, we obtain subgroups with $1{,}568$ and $859$ respondents, respectively. The distinguishing features between these two subgroups mirror the differences found between the PHNMF subgroups.  However, when we run LCA again on these two subgroups, we obtain 6 and 7 subgroups for each group, respectively.  This is again difficult to interpret.  We do not employ LPA since it requires continuous variables.

Tab.~\ref{tab:thetaresults2} depicts the cosine similarity between the coefficient vectors obtained with ordinal ridge regression on the discovered PHNMF subpopulations and those obtained from the same analysis on the entire population. 

We observe greater alignment between coefficient vectors in subpopulations obtained from earlier PHNMF splits.  However, the directions differ more as PHNMF gleans finer subpopulation structure with additional splits.  While not depicted here, we also note that the most statistically significant regression variables identified by ordinal ridge regression agree with our interpretations of the PHNMF subpopulations. For example, for Group 1.1, the variable corresponding to ``Yes, global warming is happening" is not significant since the entire group is distinguished by their agreement with this statement. However, within this subpopulation, the most significant variables relate to the extent that respondents are concerned about the impact of climate change on the future.  This coincides with the distinguishing feature of the next PHNMF split. 

\begin{table}[htbp]
\caption{Cosine similarity between coefficient vectors obtained from ordinal ridge regression on the subgroups compared with on the population as a whole. }
\centering
\begin{tabular}{lc}
\hline
\multicolumn{1}{c}{\textbf{Group Name}} & \multicolumn{1}{c}{\textbf{Cosine Similarity}}
\\
\hline \\
Group 1.2 &0.9521\\
Group 1.1.2 &0.8545\\
Group 1.1.1.2 &0.7226\\
Group 1.1.1.1.1 &0.7758\\
Group 1.1.1.1.2.1 &0.7951\\
Group 1.1.1.1.2.2 &0.7966\\
Group 2.1 &0.9983\\
Group 2.2 &0.9533\\
\hline
\end{tabular}
\label{tab:thetaresults2}
\end{table}

\section{Discussion and Future Works}
\label{sec:discussion}

Our results on synthetic data demonstrate empirically that PHNMF can identify hierarchical population structure with high accuracy.  They also demonstrate that the discovered subpopulation structure can improve downstream inference with regression methods.  Our results on real data highlight PHNMF's ability to identify interpretable and meaningful hierarchical population structure from complex data arising from real surveys. Moreover, when interpreted along with additional demographic statistics, the PHNMF splits align substantially with existing literature.

By contrast, LPA does not produce functional results for any of the synthetic or real datasets.  Additionally, the LCA splits on the real datasets are inconclusive and uninterpretable.  This empirical evidence suggests that these traditional methods for identifying latent class structure may be limited in data scenarios with complex hierarchical population structure and larger numbers of variables.

Future work may incorporate the inferential procedures within the PHNMF algorithm.  For example, rather than first performing PHNMF and then applying regression methods, one might weave methods such as semi-supervised NMF into one or more layers within the PHNMF algorithm.  Incorporation of semi-supervised NMF and other methods might provide increased flexibility for subpopulation discovery.



\bibliographystyle{IEEEtran}
\bibliography{references-submission}

\end{document}